\documentstyle[prl,aps,preprint]{revtex}

\begin{document}
\draft

\preprint{UCLA/93/TEP/47; hep-ph/9312224}
\title{$1/m_b$ Corrections in the ACCMM Model for Inclusive
Semileptonic B Decay}
\author{Grant Baillie\thanks{E-mail address: baillie@physics.ucla.edu}}
\address{Department of Physics, University of California, Los Angeles
\\ 405 Hilgard Avenue, Los Angeles, CA, 90024-1547, USA}
\date{December 5, 1993}
\maketitle
\vfill
\begin{abstract}
We re-examine the model of Altarelli, Cabibbo, Corb\`{o}, Maiani and
Martinelli for inclusive semileptonic B decay, in the light of recent
calculations in heavy quark effective theory. The model can be shown to
have no~$1/m_b$ corrections, with a suitable definition of the b quark
mass~$m_b$.  However, we find that the structure of the $1/m_b^2$~terms is
incompatible with the predictions of heavy quark effective theory. The
numerical significance of this discrepancy is discussed.
\end{abstract}
\vfill
\pagebreak

\section{Introduction}
Recently, there has been much progress in the application of heavy quark
effective theory (HQET) to the calculation of inclusive semileptonic decays
of B hadrons. It was shown by Chay, Georgi and Grinstein~\cite{Chay} that
the problem can be attacked using methods familiar from deep inelastic
scattering. In particular, one can perform an operator product expansion on
the hadronic tensor contributing to~$H_b \rightarrow X \ell
\bar{\nu_\ell}$, where $H_b$~represents a (heavy) bottom hadron, and~$X$ a
sum over final hadron states. At leading order in this expansion, the
parton model prediction results, while higher-order terms, which are
suppressed by inverse powers of the b quark mass~$m_b$, give
non-perturbative corrections. In addition, it was demonstrated the
expansion is free of $1/m_b$~terms, essentially because in the effective
theory, the operators with the correct dimensions to contribute to such
terms have matrix elements which vanish to leading order.

The $1/m_b^2$~corrections to the parton model for the inclusive
differential rate~$d\Gamma / dE_\ell$ (where~$E_\ell$ is the final lepton
energy in the B~hadron center of mass) were calculated by Bigi {\sl et
al}~\cite{Bigi}. The non-perturbative corrections enter the calculation in
the form of matrix elements of higher dimension operators: the two that
contribute in this case are
\begin{eqnarray*}
{1 \over m_b^2}
\left\langle
  H_b(v)
  \left|
     \bar{b}_v \, (v \cdot D)^2 \, b_v
  \right|
  H_b(v)
\right\rangle
\qquad
{\rm and}
\qquad
{g \over m_b^2}
\left\langle
  H_b(v)
  \left|
     \bar{b}_v \, \sigma_{\alpha\beta} \, G^{\alpha\beta} \, b_v
  \right|
  H_b(v)
\right\rangle
\end{eqnarray*}
where $H_b(v)$~denotes a B hadron with velocity~$v$, $D$~is the QCD
covariant derivative, $G_{\alpha\beta}$~is the QCD field strength tensor
and $b_v$~denotes the quark field in the heavy quark effective theory.

Subsequently, a very detailed exposition of the $1/m_b^2$~corrections was
given by Manohar and Wise~\cite{Wise}. This paper presented double
differential distributions (which also appeared around the same time in
\cite{Blok}), and also considered polarisation effects in
$\Lambda_b$~decays. In addition, it was also shown that some of the higher
order terms could be interpreted as resulting from averaging the
zeroth-order ({\sl i.e.} parton model) decay rate over the residual
4-momentum of the bound b quark inside the B hadron.

This interpretation is reminiscent of the older calculation of Altarelli
{\sl et al}~\cite{ACCMM}, which we henceforth refer to as the the ACCMM
model. In this model, bound state corrections are incorporated by
introducing a wavefunction for the two quarks inside the B meson, and then
averaging the partonic decay rate over their relative 3-momentum. The size
of the ACCMM wavefunction is determined by the Fermi momentum~$p_F$ of the
quarks inside the B meson, which is of order~$\Lambda_{\rm QCD}$.
Consequently, one can perform a large~$m_b$ expansion of the ACCMM result,
and investigate whether it can, with suitable choice of parameters, be made
to agree with the QCD predictions of~\cite{Chay,Bigi,Wise,Blok}.  This
procedure is the main focus of this letter. Throughout, we ignore the
corrections of perturbative QCD to the process, discussing bound state
corrections only. In Section~\ref{sec:HQET}, we introduce our notation and
briefly review the averaging of Manohar and Wise. The ideas behind the
ACCMM model are introduced in Section~\ref{sec:ACCMM}, where we also
attempt to find a correspondence with the newer calculation to
order~$1/m_b^2$. We present our conclusions in
Section~\ref{sec:Conclusions}.

\section{Averaging in HQET}
\label{sec:HQET}
{}From the charged-current interaction in the
Standard Model
\begin{equation}\begin{array}{rcl}
{\cal L}  & \quad = \quad &
- {\displaystyle 4 G_F \over \displaystyle\sqrt{2}} \; V_{jb} \:
   \bar{q}_j \gamma^\mu P_L b \: \:
   \bar\ell \gamma_\mu P_L \nu_\ell \quad + \quad {\rm h.c.} \\
 & \qquad \equiv \qquad &
- {\displaystyle 4 G_F \over \displaystyle \sqrt{2}} \; V_{jb} \:
   J^\mu_j \: J_{\ell\mu}
   \quad + \quad {\rm h.c.,}
\end{array}\end{equation}
(where~$j$ is a flavour index, $V$~is the CKM quark mixing matrix, $G_F$~is
the Fermi weak decay constant and $P_L$~is the projection operator~$(1 -
\gamma^5)/2$) one can derive the inclusive differential semileptonic decay
rate
\begin{equation}
{\displaystyle d\Gamma
 \over
 \displaystyle dq^2 dE_\ell dE_\nu}
\quad = \quad
{\displaystyle G_F^2 \over \displaystyle 4 \pi^3}
\, \left| V_{jb} \right|^2
\, L_{\alpha\beta} \, W^{\alpha\beta}.
\end{equation}
Here one has summed over all final states~$X_j$ containing a quark of
flavour~$j=u \hbox{ or } c$. The 4-vector~$q$ is just the sum of the
4-momenta $p$~and~$p^\prime$ of the final state lepton and
antineutrino, while $E_\ell$~and~$E_\nu$ denote the energies of these
particles. Throughout, we neglect the lepton mass~$m_\ell$.
The leptonic tensor~$L_{\alpha\beta}$ is given
by
\begin{equation}
L_{\alpha\beta} \quad = \quad
2 \left(
  p_\alpha p^\prime_\beta \; + \; p_\beta p^\prime_\alpha
  \; - \; g_{\alpha\beta} \, p \cdot p^\prime
  \; - \; i \, \epsilon_{\alpha\beta\mu\nu} p^\mu p^{\prime\nu}
\right),
\end{equation}
where our convention for the totally antisymmetric tensor
is~$\epsilon_{\rm 0123} = +1$.
$W_{\alpha\beta}$~can be expressed as
\begin{equation}
W_{\alpha\beta}
\quad = \quad
\sum_{X_j} \:
\delta^4 \left( P_{H_b} - P_X - q \right) \:
\left\langle
  H_b
  \left|
    J^{\dag}_{j\alpha} \, (0)
  \right|
  X_j
\right\rangle
\left\langle
  X_j
  \left|
    J^{\phantom{\dag}}_{j\beta} \, (0)
  \right|
  H_b
\right\rangle.
\end{equation}

In this expression, we have normalised the B hadron state~$|H_b\rangle$ as
in ref.~\cite{Wise} to remove extra factors of mass from the decay
rate~(2). In the parton model, $W_{\alpha\beta}$~is approximated by its
value from free quark decay:
\begin{equation}\begin{array}{rcl}
W_{\alpha\beta}
& \quad = \quad &
W^{(0)}_{\alpha\beta} \, \left( m_b, v \right)
\\
& \quad = \quad &
\delta
\left(
   m_b^2 - 2 m_b \left(E_\ell + E_\nu \right) + q^2 - m_j^2
\right)
\\
 & &
 \times \left\{
  \, - {1 \over 2} g_{\alpha\beta} \left( m_b - E_\ell - E_\nu \right)
  \: + \: m_b \, v_\alpha v_\beta
  \: - \: {1 \over 2} i \epsilon_{\alpha\beta\mu\nu} v^\mu q^\nu
 \right\}.
\end{array}
\end{equation}
Here we have introduced the velocity~$v_\mu$ of~$H_b$; in addition we
explicitly note the dependence on $m_b$~and~$v_\mu$ because these will
later be averaged over.  As shown in reference~\cite{Wise}, most of the
$1/m_b^2$~corrections to the parton model result to~$W_{\alpha\beta}$ can
be obtained by averaging the RHS of equation~(5) over the residual
4-momentum~$k_\alpha$ of the bound b~quark. If one defines the
mass~$m_b^\prime$, velocity~$v_\mu^\prime$ and momentum~$P_\mu$ of the
moving b~quark via
\begin{equation}
P_\mu \quad = \quad
m^\prime_b v^\prime_\mu \quad = \quad
m_b v_\mu \; + \; k_\mu \: {\rm ;} \quad
\left( v^{\prime 2} = v^2 = 1 \right),
\end{equation}
then one can expand its contribution to~$W_{\alpha\beta}$ to
second order in~$k$:
\begin{equation}
{\displaystyle 1 \over \displaystyle v^\prime_0}
W^{(0)}_{\alpha\beta} \left( m^\prime_b, v^\prime \right)
\quad \approx \quad
\left.
  \left(
      1
      \; + \;
      k^\mu {\displaystyle \partial \over \displaystyle \partial k^\mu}
      \; + \;
      {1 \over 2} k^\mu k^\nu
      {\displaystyle \partial^2 \over
       \displaystyle \partial k^\mu \partial k^\nu}
  \right)
  \, {\displaystyle 1 \over \displaystyle v^\prime_0}
  W^{(0)}_{\alpha\beta} \left( m^\prime_b, v^\prime \right)
\right|_{k=0}.
\end{equation}
Note that this expression contains an expansion of the $\delta$~function
in~(5), which cannot be justified everywhere in the Dalitz plot. In fact,
the OPE contains an analagous expansion (of the $j$-quark propagator),
which breaks down in some of the regions where a partonic description is
not reasonable~\cite{ISGW}. We will assume throughout this paper that we
are in a region of phase space where this truncation is valid.

It turns out that almost all the terms coming from the OPE analysis up to
order~$1/m_b^2$ can be reproduced by replacing~$W_{\alpha\beta}$ by the
average (over~$k$) of equation~(7), with the average values
\nopagebreak\begin{mathletters}
\begin{eqnarray}
\left\langle k^\alpha \right\rangle
& \quad = \quad &
E_b m_b v^\alpha
\\
\left\langle k^\alpha k^\beta \right\rangle
& \quad = \quad &
-{2 \over 3} \, m_b^2 \, \left( g^{\alpha\beta} - v^\alpha v^\beta \right).
\end{eqnarray}
\end{mathletters}

{}From this point of view, the order of magnitude of the quantities
$E_b$~and~$K_b$ might not be obvious; however, they can be expressed as
matrix elements of operators between B meson states. As a result, they are
expected to be of order~$(\Lambda_{\rm QCD}/m_b)^2$. Incidentally, one
should note that the corrections which cannot be accounted for via this
averaging are proportional to the parameter~$G_b$, which is defined in HQET
by
\begin{equation}
G_b \quad = \quad
{\displaystyle Z_b \over 4 m_b^2} \,
\left\langle
  H_b \left( v,s \right)
  \left|
    \: \bar{b}_v \, g G_{\alpha\beta} \; \sigma^{\alpha\beta} \, b_v \:
  \right|
  H_b \left( v,s \right)
\right\rangle,
\end{equation}
 where $s$~and~$v$ denote the spin and velocity of the
heavy meson~$H_b$, and~$Z_b$ is a renormalisation constant. To leading order
in~$1/m_b$, this operator is a spin-spin interaction between the heavy
quark and its surrounding ``brown muck''. A simple physical
interpretation of its effects has not yet been given, however.

\section{Comparison with ACCMM}
\label{sec:ACCMM}
The ACCMM model introduces bound state
corrections to the parton result for B~meson decay by considering the
meson to be made up of a b and spectator quarks. Furthermore, the
energy of the spectator is assumed to be given in terms of its
3-momentum~${\bf p}_{sp}$ by
\begin{equation}
E_{sp} \quad = \quad \sqrt{\displaystyle {\bf p}^2_{sp} + m_{sp}^2}.
\end{equation}
where~$m_{sp}$ is a free parameter of order~$\Lambda_{\rm QCD}$. The
4-momentum~$P^\mu$ of the b quark is now fixed by the requirement that the
sum of the b and spectator momenta be that of the B~hadron. One finds
that\begin{equation}\begin{array}{rcl}
P_0 & \quad = \quad &
M_B \; - \; E_{sp} \quad = \quad
M_B - \sqrt{\displaystyle {\bf p}^2_{sp} + m_{sp}^2}
\\
{\bf P} & \quad = \quad & -{\bf p}_{sp}
\end{array}\end{equation}
and hence that the square of the invariant mass
of the moving b quark is just
\nopagebreak\begin{equation}
m_b^{\prime 2} \quad = \quad P^2 \quad = \quad
M_B^2 + m_{sp}^2 - 2 M_B \sqrt{\displaystyle {\bf p}^2_{sp} + m_{sp}^2}.
\end{equation}

To perform the averaging over the motion of the quarks inside the hadron,
one introduces a wavefunction~$\psi({\bf p}_{sp})$. Typically, one expects
that this function will have a width of order~$\Lambda_{\rm QCD}$ and that
it will fall off rapidly for values of momentum~${\bf p}_{sp}$ larger than
this value. Also, because we understand the B~meson to be an L=0~state, we
assume that~$\psi({\bf p}_{sp})$ is spherically symmetric. In practice, one
takes
\begin{equation}
\left| \psi( {\bf p}_{sp} ) \right|^2
\quad = \quad
N \; \exp \, \left( {\bf p}_{sp}^2 / p_F^2 \right),
\end{equation}
where~$N$ is a normalisation factor and~$p_F$, the Fermi momentum of the
quarks inside the hadron, is taken to be an adjustable parameter of
order~$\Lambda_{\rm QCD}$. One can now write down the hadronic tensor in
the model, by assuming it is just the average over the quark~3-momentum of
the boosted free quark value:
\begin{equation}
W_{\alpha\beta}
\quad = \quad
\int d^3 {\bf p}_{sp}
\left| \psi( {\bf p}_{sp} ) \right|^2 \;
{\displaystyle 1 \over \displaystyle v^\prime_0}
W^{(0)}_{\alpha\beta} \left( m^\prime_b, v^\prime = P / m^\prime_b \right).
\end{equation}
In this expression, we have explicitly included a Lorentz time-dilation
factor in the integral, which was omitted in~\cite{ACCMM}. It should also
be noted that in \cite{ACCMM} the integral is cut off at the maximum value
of $|{\bf p}_{sp}|$~allowed in the decay; however, the wavefunction is so
small at this point (the exponential factor in~(13) is tiny there) that we
can ignore the effect of this upper limit in what follows.

Since the wavefunction factor in~(14) effectively restricts~$|{\bf
p}_{sp}|$ to be at most of order~$p_F$, we can envisage expanding the
factor~$W^{(0)}_{\alpha\beta}(m_b^\prime, v^\prime) / v^\prime_0$ in~(14)
by treating~${\bf p}_{sp}$, $p_F$~and~$m_{sp}$ as small. Clearly, the first
term in this expansion will just be~$W^{(0)}_{\alpha\beta}(M_B, v)$. The
fact that this depends on~$M_B$ and not~$m_b$ is not accidental: the ACCMM
model was specifically constructed to try to circumvent the dependence of
the partonic decay rate on the unknown parameter~$m_b$. Just as was the
case in HQET, our expansion will contain average values, which are now
given (as averages over~${\bf p}_{sp}$) by
\begin{equation}
\left\langle f \right\rangle
\quad = \quad
\int d^3 {\bf p}_{sp}
\left| \psi( {\bf p}_{sp} ) \right|^2 \;
f({\bf p}_{sp}).
\end{equation}

If the two types of averaging were to produce identical results to this
order, we would be able to express the HQET parameters~$m_b$,
$E_b$~and~$K_b$ in terms of~$M_B$, $p_F$~and~$m_{sp}$. In fact, there would
be some dependence on the ACCMM wavefunction through, for example, a
quantity like~$\langle{\bf p}_{sp}\rangle$. Because our expansion starts
at~$M_B$, and because the HQET expansion contains no $1/m_b$~terms, we will
evidently have to absorb $1/M_B$~terms into the definition of the quark
mass~$m_b$.  Evidently, in ACCMM we have from~(11)
\begin{equation}
\left\langle P^\alpha \right\rangle
\quad = \quad
v^\alpha \;
\left(
  M_B \, - \,
  \left\langle
     \sqrt{\displaystyle {\bf p}^2_{sp} + m_{sp}^2}
  \right\rangle
\right).
\end{equation}
Consistency with (6)~and~(8a) can only be
obtained if
\begin{equation}
m_b \, (1 + E_b)
\quad = \quad
M_B \, - \,
\left\langle \sqrt{\displaystyle {\bf p}^2_{sp} + m_{sp}^2} \right\rangle.
\end{equation}
The 4-momentum~$k$ in~(6) can now be expressed in
the B rest frame as
\begin{equation}
k^\mu \quad = \quad
\left(
  E_b M_B \; - \;
  \sqrt{\displaystyle {\bf p}^2_{sp} + m_{sp}^2} \; + \;
 \left\langle
     \sqrt{\displaystyle {\bf p}^2_{sp} + m_{sp}^2}
  \right\rangle,
  \: - {\bf p}_{sp}
\right),
\end{equation}
so that we have, to leading nonvanishing order
\begin{equation}\begin{array}{rcl}
\left\langle k^\alpha k^\beta \right\rangle
& \quad = \quad &
\left(
  \left\langle
    {\bf p}^2_{sp} + m_{sp}^2
  \right\rangle
  -
  \left\langle
    \sqrt{\displaystyle {\bf p}^2_{sp} + m_{sp}^2}
  \right\rangle^2
\right)
v^\alpha v^\beta
\\
& &
- {1 \over 3}
\left\langle {\bf p}^2_{sp} \right\rangle
\left(
  g^{\alpha\beta} - v^\alpha v^\beta
\right).
\end{array}\end{equation}
On the RHS of this equation, the first term comes
from the $00$-component in the B rest frame, while the second is the
$ij$-th, taking into account rotational symmetry. The $0i$- and
$i0$-components can easily be seen to vanish.

Since the two tensors
in~(19) are linearly independent, the only way to reconcile this
equation with~(8b) is to have
\begin{equation}
K_b \quad = \quad
{\displaystyle
\left\langle
{\bf p}^2_{sp}
\right\rangle
\over \displaystyle 2 m_b^2}
\end{equation}
and also for the coefficient of~$v^\alpha v^\beta$ to be of
order~$\Lambda_{\rm QCD}^3 / m_b$, {\sl i.e.} to vanish to the order we are
considering. However, for a general wavefunction of size~$p_F$, this
quantity will be of order~$p_F^2$.  In addition, one can compute the
contribution of the extra term to the quantity~$L_{\alpha\beta}
W^{\alpha\beta}$ in~(2), to make sure it does not vanish. We find a
contribution to this quantity of
\begin{equation}\begin{array}{l}
\begin{array}{rl}
2 A_b m_b^2 E_\nu
&
\left\{
  \delta^\prime (X) \:
  \left(
    6 m_b E_\ell - 4 E_\ell^2 - 4 E_\ell E_\nu - q^2
  \right)
\right.
\\
&
+
\left.
  2 \; \delta^{\prime\prime} (X) \:
  \left(
    m_b - E_\ell - E_\nu
  \right)^2
  \left(
   2 m_b E_\ell - q^2
  \right)
\right\}
\end{array}
\\
  \quad
  \hbox{\rm with }
  X \equiv
  m_b^2 - 2 m_b \left( E_\ell + E_\nu \right) + q^2 - m_j^2,
  \quad
\end{array}
\end{equation}
where we have defined the dimensionless parameter~$A_b$ by
\begin{equation}
A_b \, m_b^2
\quad = \quad
\left\langle
  {\bf p}^2_{sp} + m_{sp}^2
\right\rangle
-
\left\langle
  \sqrt{\displaystyle {\bf p}^2_{sp} + m_{sp}^2}
\right\rangle^2.
\end{equation}
Incidentally, one should note that this contribution cannot just cancel the
Lorentz time-dilation factor mentioned after~(14), as that gives an overall
factor~$(1 - E_b)$ to the decay.  Consequently, it seems to be impossible
to find an exact correspondence between the averaging of the
phenomenological ACCMM model and the QCD calculations of HQET.

\section{Conclusions}
\label{sec:Conclusions}

We have analysed, in two different approaches, the corrections to the
parton model for inclusive semileptonic B meson decay up to order
$1/m_b^2$. Although we have found that the ACCMM model cannot be put into
exact correspondence with the HQET calculation, we should point out that
the discrepancy between the two is likely to be small numerically. This is
mainly due to the fact that $1/m_b$ corrections vanish in both cases if
equation~(17) is satisfied. As demonstrated in~\cite{Wise}, the
$1/m_b^2$~terms adjust the parton result by around a percent in the region
where the expansion makes sense. Also, the additional term appearing in
$\langle k^\alpha k^\beta \rangle$ (cf. equation (19)) in the ACCMM case is
likely to be suppressed compared to the other term for a monotonically
decreasing wavefunction. This term would dominate the other corrections in
the $m \rightarrow \infty$~limit, but for the B it is practically
indistinguishable from the higher order terms we have ignored.

The ACCMM model's disagreement can be traced to its assumption that the
spectator quark in the B meson can be treated as an on-shell particle of
mass~$m_{sp}$, of the order of the QCD scale. This led to one having to
define $m_b$ in such a way so as to avoid $1/m_b$ terms in the expansion.
However, making this definition was not sufficient to ensure that the
$1/m_b^2$~terms had the correct form. In the context of HQET, it is clear
that the ``spectator quark'' of ACCMM corresponds to ``brown muck''.
However, it seems unlikely that the latter can be treated simply as an
object of mass $m_{sp}$. If one wanted to improve the model
to make it compatible with HQET, one could start by relaxing the on-shell
condition for the spectator quark.

Finally, it is worth noting that equations (6)~and~(8), together with~(18)
and the way we defined the quark mass in ACCMM, suggest that, as far as
these inclusive processes are concerned, the parameter~$E_b$ can be
absorbed into a redefinition of $m_b$. This can be confirmed by examining
the expressions in section~6 of reference~\cite{Wise}: the $E_b$~correction
terms have the form
\begin{eqnarray*}
E_b m_b {\displaystyle \partial \over \partial m_b}
\,
\hbox{(parton model quantity)},
\end{eqnarray*}
where the partial derivative is taken at fixed lepton momenta. Presumably,
this should still hold true in the case of $\tau$~decay modes, where the
lepton masses cannot be ignored~\cite{Koryakh}. This observation might be
useful if one were trying to fit the HQET parameters to an experimental
spectrum, for example.

\section*{Acknowledgements}

This work was supported by United States Department of Energy Grant
\hbox{AT03--88ER~40384}, Task~C. The author would like to thank
Roberto~Peccei for suggesting this project, and Dario~Zappal\`{a},
Kingston~Wang, Brian~Hill, Antonio~Bouzas, and Mark~Wise for useful
discussions.

\end{document}